\documentclass{aa}
\usepackage{graphicx}
\usepackage{txfonts}
\begin{document}
\title{An estimate of the structural parameters of the Large Magellanic Cloud using red clump stars}
\author{Smitha Subramanian\inst{1,2}, Annapurni Subramaniam\inst{1}}
\institute{Indian Institute of Astrophysics, Koramangala II Block, Bangalore-560034, India\\
	   Department of Physics, Calicut University, Calicut, Kerala\\
           \email{smitha@iiap.res.in, purni@iiap.res.in}}
\date{Received, accepted}
\abstract
{The structural parameters of the disk of the Large Magellanic Cloud (LMC) are estimated.}
{We used the red clump stars from the {{\it VI}} photometric data of the Optical 
Gravitational Lensing Experiment (OGLE III) survey and from the Magellanic 
Cloud Photometric Survey (MCPS) to estimate the structural parameters of the
 LMC disk, such as the inclination, $i$ and the position angle of the
 line of nodes (PA$_{lon}$), $\phi$.} 
 {The observed disk region is divided into sub-regions. 
The dereddened peak I magnitude of the red clump stars in each sub-region 
is used to obtain the relative distances and hence the z coordinate. 
The RA and Dec of each sub-region is converted into x and y cartesian 
coordinates. 
A weighted least-square plane-fitting method is applied to 
these x,y,z data to estimate the structural parameters of the LMC disk.} 
{We find an inclination of $i$ =23$^o$.0$\pm$0$^o$.8 and PA$_{lon}$, $\phi$ =
 163$^o$.7$\pm$1$^o$.5 for the LMC disk using 
the OGLE III data and an inclination of $i$=37$^o$.4$\pm$2$^o$.3 and PA$_{lon}$ 
$\phi$ = 141$^o$.2$\pm$3$^o$.7 for the LMC disk using 
the MCPS data. Extra-planar features, which are in front 
as well as behind the fitted plane, are seen in both the data sets.}
{Our estimates of the inclination and position angle of the line of nodes
 are comparable with some of the previous estimates. The effect of choice of center,
reddening, and area covered on the estimated parameters are discussed. 
Regions in the northwest, southwest and southeast of the LMC disk are warped
with respect to the fitted plane. 
We also identify a symmetric but off-centered warp in the inner LMC. 
We identify that the structure of the LMC 
disk inside the 3 degree radius is different from the outside disk in a way that the inner LMC has
relatively less inclination and relatively large PA$_{lon}$. 
The 3D plot of the LMC disk suggests an off-centered increase in the 
inclination for the northeastern regions, which  might be due to tidal effects. We suggest that the
variation in the planar parameters estimated by various authors as well as in this study is caused by the difference in
coverage and the complicated inner structure of the LMC disk. In the inner LMC, the stellar and the HI
disk are found to have similar properties.
}
\keywords{(galaxies:) Magellanic Clouds;
galaxies: structure;
stars: horizontal-branch
}
\authorrunning{Subramanian \& Subramaniam}
\titlerunning{Structure of the Large Magellanic Cloud}
\maketitle

\section{Introduction}
The Large Magellanic Cloud (LMC) is one of our nearest neighbors and is located 
at a distance of around 50 kpc. Magellanic Clouds (MCs) were believed 
to have had interactions with our Galaxy as well as between each other
(\cite{wes97}). It is also believed that the tidal forces due to 
these interactions have caused structural changes in this galaxy. 
The study of the LMC's structure is important to understand the effect 
of interactions in this galaxy.\\

The LMC is believed to be a disk galaxy with planar geometry, and the orientation 
measurements of the LMC disk plane have been done previously by various 
authors using different tracers.
By assuming the LMC disk to be circular when viewed face-on, 
\cite{df72} found a disk inclination of 
$i$ = 27$^o$$\pm$2$^o$and a position angle of
 the line of nodes (PA$_{lon}$), $\phi$ = 170$^o$$\pm$ 5$^o$ from the elliptical
 outer isophotes in red exposures.
Based on the analysis of spatial variations 
in the apparent magnitude of asymptotic giant branch (AGB) stars in the 
near-IR colour-magnitude diagrams extracted 
from the Deep Near-Infrared Southern Sky Survey (DENIS) and Two Micron 
All-Sky Survey (2MASS), \cite{vc01} estimated an  $i$ = 34$^o$.7$\pm$6$^o$.2
and $\phi$ = 122$^o$.5$\pm$8$^o$.3 for the 
LMC disk between 2$^o$.5 to 6$^o$.7 from the LMC center. \cite{os02}
 derived an $i$ = 35$^o$.8 $\pm$2$^o$.4 and $\phi$ = 145$^o$ $\pm$ 4$^o$ by 
studying the red clump (RC) magnitudes in the inner LMC, excluding the bar region. 
They also showed that the southwestern part of the LMC disk is warped. 
Recently, \cite{k09} 
derived an $i$ = 23$^o$.5 $\pm$ 0$^o$.4 and $\phi$ = 154$^o$.6 $\pm$ 1$^o$.2,
using the {\it JH} photometric data of RC stars 
from the Infrared Survey Facility (IRSF)  Magellanic Clouds Point Source Catalogue.
Cepheids are also used  to obtain the orientation measurements of the 
LMC disk. \cite{cc86} analyzed a sample of 73 cepheids and 
obtained $i$ = 28$^o$.6 $\pm$ 5$^o$.9 and $\phi$ = 142$^o$.4$\pm$7$^o$.7. 
\cite{n04} based on a sample of 2000 MACHO cepheids obtained an 
$i$ = 30$^o$.7 $\pm$ 1$^o$.1 and $\phi$ = 151$^o$$\pm$2$^o$.4.
\cite{p04} obtained an $i$ = 27$^o$ $\pm$ 6$^o$ and $\phi$ = 127$^o$
 $\pm$ 10$^o$ from the analysis of 92 near infrared light curves of Cepheids.
Various studies of HI gas estimated the structural parameters of the 
HI disk of the LMC. \cite{F77} derived an inclination of 33$^o$$\pm$3$^o$ and PA$_{lon}$ of 
168$^o$$\pm$4$^o$ by geometrical means. The HI velocity studies by \cite{LR92} 
revealed two kinematic components, the L (lower velocity) component and the
 D (disk) componenet. The PA$_{lon}$ of around 162$^o$ was
estimated for the disk component. \cite{k98} estimated the PA$_{lon}$ of HI disk
to be around 168$^o$ and an inclination of 22$^o$ $\pm$ 6$^o$. \\  

The red Clump (RC) stars are core helium burning stars, which are the metal-rich 
and slightly more massive counter parts of the horizontal branch stars. 
They have a tightly defined color and magnitude, and appear as an
easily identifiable component in the color-magnitude diagrams (CMDs). 
RC stars were used as standard candles for distance determination by 
Stanek et al. (1998). They used the intrinsic luminosity to determine 
the distance to the LMC. 
\cite{s03} used the constant magnitude of RC stars to show that 
the LMC has structures and warps in the bar region.
Their characteristic color was used by \cite{s05} to
estimate the reddening map towards the central LMC. 
\cite{SS09} estimated the depth of the Magellanic Clouds using 
the dispersions in the magnitude and color distributions of RC stars.
As mentioned in the earlier paragraph, RC stars are also used to estimate the 
orientation measurements of the LMC disk plane.\\
 
In this paper we use the photometric data of the RC stars in the {\it V} and {\it I} pass bands
from the Magellanic Cloud Photometric Survey (MCPS) and the Optical 
Gravitational Lensing Experiment (OGLE III) to estimate the 
structural parameters of the LMC disk plane. These catalogs have homogeneous and continuous sampling of stars
spread over the inner LMC up to a radius of about 6 degrees. 
\cite{os02} studied the structure of the LMC using RC stars in discrete and widely separated pointings
located away from the bar region. \cite{k09} used 
{\it JH} photometric data of RC stars in the inner region of the LMC to study the 
structure, where the sample has contamination from AGB stars. 
Compared to the above two studies, we use a homogeneous and continuous sample of RC stars, with minimal
contamination from stars in other evolutionary stages.\\

The plan of this paper is as follows. Data sources are explained in 
the next section. In Sect.3 we give details of the analysis. 
Results are given in Sect.4 and their implications are discussed in 
Sect.5, followed by conclusions in Sect.6.\\    

\section{Data}
The OGLE III survey (\cite{u08}) presented {\it VI}
photometry of 40 deg$^2$ of the LMC consisting of about 35 million
stars. We divided the observed region into 1854 regions (with
a reasonable number of RC stars, 500 - 9000) with a bin size of
8.88 x 8.88 arcmin$^2$. 
Regions with RC stars in the range 500 - 700 are located 
in the eastern and western ends of the disk region
covered by the OGLE III. 
The average photometric error of RC stars
in I and V bands are around 0.05 mag. Photometric data with errors
less than 0.15 mag are considered for the analysis. 
For each sub-region the (V$-$I) vs I CMD is 
plotted and RC stars are identified. A sample CMD is shown in the 
upper left panel of figure 1. For all the regions, RC stars 
are well within the box of CMD, with boundaries 0.65 - 1.35 mag in 
(V$-$I) colour and 17.5 - 19.5 mag in I magnitude.\\

The Magellanic Cloud Photometric Survey (MCPS, \cite{z04}) 
 of the central 64 square degrees of the LMC obtained
photometric data of around 24 million stars in the {\it U, B, V,} and {\it I}
pass bands. Data with errors in the {\it V} and {\it I} pass bands less than
 0.15 mag are taken for the
analysis. The regions away from the bar are less dense compared
to the bar region. The total observed regions are divided into
1512 sub-regions each with an area of approximately 10.53 x 15 
arcmin$^2$. Out of 1512 regions only 1377 regions have
a reasonable number of RC stars (100-2000) to do the analysis. (V$-$I) vs.
I CMDs for each region are plotted, and RC stars are
identified as described above. 

\section{Analysis}
The dereddened peak RC magnitude is used to obtain the 
structural parameters
of the LMC disk. RC stars occupy a compact region in the CMDs and they
have a constant characteristic I band magnitude and  (V$-$I) color. Their
number distribution profiles resemble a Gaussian. 
The peak values of their color and magnitude distributions are used to obtain the 
dereddened RC magnitude and hence the structural parameters of the LMC disk. 
The analysis is similar to that done by \cite{os02}.\\

To obtain the number distribution of the
RC stars, they are binned in both color and magnitude with a bin
size of 0.01 and 0.025 mag, respectively. These distributions are
fitted with a Gaussian + Quadratic polynomial. The Gaussian represents the
RC stars and the other terms represent the red giants in
the region. A nonlinear least-square method is used for fitting 
and the parameters are obtained. In the lower two panels of Fig.1, 
the distributions as well as the fitted curves 
for an LMC OGLE III region are shown.  The left lower panel is the  
color distribution and right panel is the magnitude distribution. 
The parameters obtained are the coefficients of
each term in the function used to fit the profile, error in the
estimate of each parameter, and reduced $\chi$$^2$ value. For both data sets we estimated the peaks in I mag and (V$-$I) mag of the distributions, 
associated errors with the parameters, and goodness of fit.\\ 

The reduced $\chi$$^2$ values and the fit-error of the peak of 
the I magnitude distribution are plotted against RA in the right middle 
panel and right upper panel of Fig.1 respectively.
Regions with peak errors greater than 0.1 mag and those with reduced  
$\chi$$^2$ value greater than 2.0 are omitted from the analysis. 
Thus the regions used for final analysis became 1262 for the OGLE III data 
and 1231 for the MCPS data.\\

\begin{figure}
\resizebox{\hsize}{!}{\includegraphics{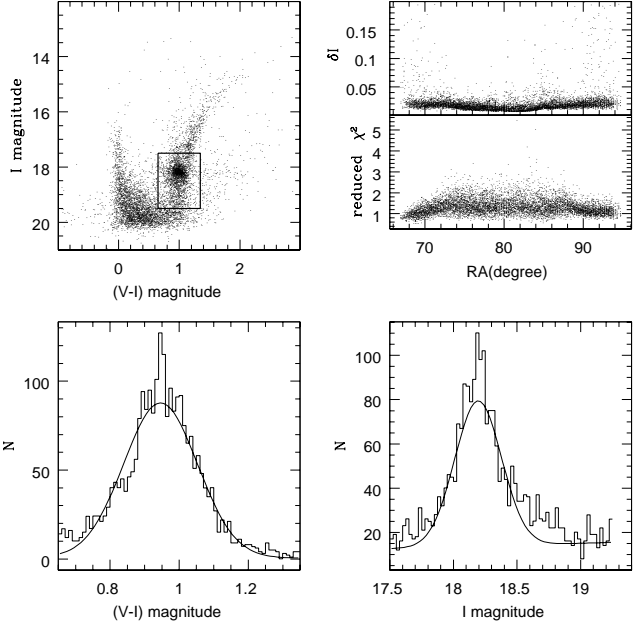}}
\caption{Color magnitude diagram of an LMC OGLE III region is shown in the 
upper left panel. The box used to identify the RC population is also shown.
Typical color (lower left panel) and magnitude (lower right panel) 
distribution of Red Clump stars are shown. 
The best fit to the distributions are also shown. The reduced $\chi$$^2$  
of the magnitude distribution and the fit error of the peak I magnitude 
as a function of RA are shown in the right middle panel and right top 
panel respectively.}
\end{figure}

The peak values of the color, (V$-$I) mag at each location
is used to estimate the reddening. The reddening is calculated
using the relation {E(V$-I$) = (V$-$I)$_{obs}$ $-$ 0.92 mag.
 The intrinsic color of the RC stars is taken to be 0.92 mag (\cite{os02}.
 The reddening values were found to be  
negative for 451 locations in the MCPS data set.
These regions were mostly located near the center. 
 \cite{z04} found many regions with negative A$_v$ values 
while estimating the extinction towards the LMC using the MCPS data and 
assigned an extinction of zero for those regions. They suggested 
that the negative extinction values are due to observational uncertainties. 
Reddening plays an important role in the estimation of peak magnitude and hence the estimated structure, thus the regions 
which showed negative values for reddening estimates 
are omitted from our analysis. Hence the number
of regions used for final analysis became 780 for the MCPS data. 
Assigning a zero reddening to those regions with negative reddening estimates 
and its impact on the estimates of the LMC disk parameters
are discussed in Sect.5. 
The selection of the  
intrinsic (V$-$I) color of the RC stars may also be the 
reason for obtaining negative reddening values. For the MCPS data, 
the intrinsic (V$-$I) color of the RC stars can be set to produce the 
median reddening obtained by \cite{sch98} towards the LMC.  
Implications of resetting the intrinsic (V$-$I) color of RC stars
for the MCPS data are discussed in detail in Sect.5. 
Negative reddening was not found for any regions in the OGLE III data.\\ 
 
The interstellar extinction is estimated by 
A$_I$=1.4xE(V$-$I) (\cite{os02}). After correcting the mean
I mag for interstellar extinction, I$_0$ for each region is estimated.
The difference in I$_0$ between regions is a measure of the relative
distances, so that $∼$ 0.1 mag in $\Delta$I corresponds to 2.3 kpc
in distance. The variation in I$_0$ mag is converted into relative distances. 
 The relative distance is \\
 
$\Delta$D = (I$_0$ mean $-$ I$_0$ of each region)x23 kpc.\\

The error in I$_0$ is estimated as\\ 

$\delta$$I_0$$^2$ = (avg error in peak I)$^2$ + (avg err in peak (V-I))$^2$ .\\

Here, the variation in I$_0$ is considered only due to
the line of sight distance variation within the galaxy.
The error in magnitude is also converted into error in 
distance.\\

In this analysis, we have not incorporated the incompleteness
due to crowding, especially in the central regions where the
effect is expected to be prominent. In order to estimate the
effect due to crowding and the incompleteness, we compared
the I$_0$ values with and without 
incompleteness correction (\cite{SS09}). \cite{SS09}
 used the OGLE II data for the analysis.
 They did not find any significant difference between the
 parameters, suggesting that the incompleteness/crowding 
does not affect the results presented here.
\cite{ss09apj} also compared the I$_0$ 
values obtained from the OGLE III with the I$_0$ values obtained from the OGLE II, 
where incompleteness correction was incorporated. They did not find
any significant change in the I$_0$ values.\\


The relative distance of each region of the LMC is obtained from 
the variation in the I$_0$ magnitude. Then the x,y, and  z coordinates are obtained using the transformation 
equations given below (\cite{vc01}, see also Appendix A of \cite{wn01}).\\\\
  x = -Dsin($\alpha$ - $\alpha_0$)cos$\delta$,\\\\
  y = Dsin$\delta$cos$\delta_0$ - Dsin$\delta_0$cos($\alpha$ - $\alpha_0$)cos$\delta$,\\\\
  z = D$_0$ - Dsin$\delta$sin$\delta_0$ - Dcos$\delta_0$cos($\alpha$ - $\alpha_0$)cos$\delta$,\\

 where D$_0$ is the distance to the center of the LMC and D, the distance to the each sub-region is given by D = D$_0$ + $\Delta$D.
The ($\alpha$, $\delta$) and ($\alpha_0$ , $\delta_0$) represents the RA and Dec of the 
  region and the center of the LMC respectively. 
In our analysis, the optical center of the LMC, $05^h19^m38^s.0$ $-69^o27'5".2$ (J2000) 
(\cite{df72}) is taken as the center of the LMC.\\ 

Once we have the x,y, and  z coordinates we can apply a weighted least square
plane fit to obtain the structural parameters of the LMC disk. 
The equation of the plane used for the plane fit is given by \\

     Ax+By+Cz+D=0, with the constraint A$^2$ + B$^2$ + C$^2$ = 1.\\

From the coefficients of the plane A,B \&C, the inclination, $i$ and the position 
angle of line of nodes (PA$_{lon}$), $\phi$ can be calculated using the formula given below.\\

    Inclination, $i$ = arccos(C) \\
    
    PA$_{lon}$, $\phi$ = arctan(-A/B)-sign(B)$\pi$/2.\\
  
A weighted least-square plane fit is applied to the x,y, and  z data 
of the MCPS \& OGLE III data sets. The inclination,$i$ and the  
PA$_{lon}$ of the LMC disk are estimated 
from the plane-fitting procedure.\\ 

We calculated the deviations of the LMC disk 
from the plane with estimated coefficients. 
The expected z for a plane is calculated with the 
equation of a plane, Ax+By+Cz+D =0. The difference in the expected and 
calculated z values is taken as the deviation of the LMC disk from the 
plane. Thus the extra-planar features of the LMC disk are identified 
and quantified. Once the deviations are estimated, the regions with 
deviations above three times the error in z are omitted and the plane-fitting 
procedure is applied to the remaining regions to re-estimate the 
structural parameters of the LMC disk plane.\\

The error in the estimate of the LMC disk parameters is calculated from the 
error associated with the z values. The plane-fitting procedure 
is repeated with the positive and negative deviation of z values 
and the LMC disk parameters are calculated again. Thus the range of 
structural parameters are estimated, which are converted as the error 
in the estimate of the parameters.\\

\begin{figure}
\resizebox{\hsize}{!}{\includegraphics{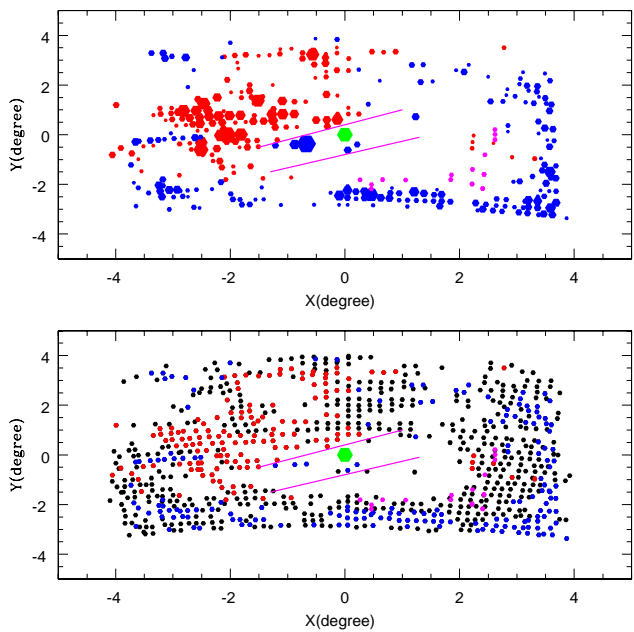}}
\caption{In the lower panel, black dots represent the regions on the fitted
 plane, red dots represent the regions behind the fitted plane and 
the blue dots represent the regions which are in front of the fitted 
plane. The upper panel shows only the regions with deviations greater than 
3 sigma from the LMC disk plane. The size of the points are proportional 
to the amplitude of the deviations. Magenta dots in both panels are 
the regions which are suggested as warps by \cite{os02}.
The green hexagon in both plots represents the optical center of the 
LMC.}
\end{figure}
\section{Results}
The structural parameters of the LMC disk are estimated using the 
dereddened mean magnitude of RC stars. The MCPS and OGLE III data 
sets are used for the estimate. A weighted plane-fitting procedure 
applied to 780 regions of the MCPS data set gives an inclination 
of $i$= 38$^o$.2$\pm$5$^o$.0 \& $PA_{lon}$, 
$\phi$ = 141$^o$.5$\pm$9$^0$.8 for the LMC disk.
The deviation of the LMC disk regions from the estimated plane 
are calculated as explained in the previous section. 
The average error in the estimate of I$_0$ mag is 
converted into distance and its around 500 pc for the MCPS data. 
Deviations above 3 sigma are considered as significant deviations from 
the fitted plane. 
Fig.2 shows the deviation of the MCPS regions from the plane.
In the bottom panel, all the regions used 
for the analysis are plotted. The black points are those which are on the 
fitted plane, red points are disk regions which are behind 
the plane and blue points are the disk regions which are 
in front of the plane. In the upper panel only the regions 
with deviations above 3 sigma are plotted and the size of the points 
are proportional to the amplitude of the deviation. Here also, red points 
are regions behind the fitted plane and blue are in front of the fitted plane. 
From the plots we can see that there are many regions which deviate 
from the planar structure of the LMC disk. 
The plot shows that the RC stars in the regions southeast, 
southwest and  northwest of the LMC disk are brighter than what is expected from the plane fit. Also, some regions northeast to the LMC bar are dimmer than expected.\\

The RC stars in the LMC disk are a heterogeneous population, and therefore  
they would have a range in mass, age, and metallicity. The density of stars in 
various locations will also vary with the local star-formation rate as a 
function of time. These factors result in a range of magnitude and color of 
the net population of RC stars in any given location and would contribute to 
the observed peaks in magnitude and color distributions. Therefore, the deviations 
found in some regions
may also be due to these population differences of RC stars 
Then the brightening of RC stars in the southeast, 
southwest and northwest of the LMC disk indicates either a different 
RC population and/or these regions are warped. 
Similarly, the dimming of RC stars in the north east of the LMC 
bar indicates a difference in RC population and/or 
these regions are behind the fitted plane.
Based on the studies of the LMC clusters, \cite{G06} found that the LMC 
lacks the metallicity gradient typically seen in the galaxies. 
Studies by \cite{sa02}, \cite{os02} and \cite{vc01} found no noticeable 
change in age and metallicity of the RC population in the 
central region of the LMC. 
As the regions we study are located in the central region of the LMC, 
the effects of population difference of RC stars in these regions  
are likely to be negligible. 
Hence we can see from these plots that 
the southeast, southwest and the northwestern regions 
of the LMC disk may be warped and the regions northeast to the bar of 
the LMC are behind the fitted plane. In the figure, the regions 
suggested by \cite{os02} as warps are over plotted as 
magenta points. Their points in the southwestern regions are 
near the warped regions suggested by us, though they do not coincide.\\
 
\begin{figure}
\resizebox{\hsize}{!}{\includegraphics{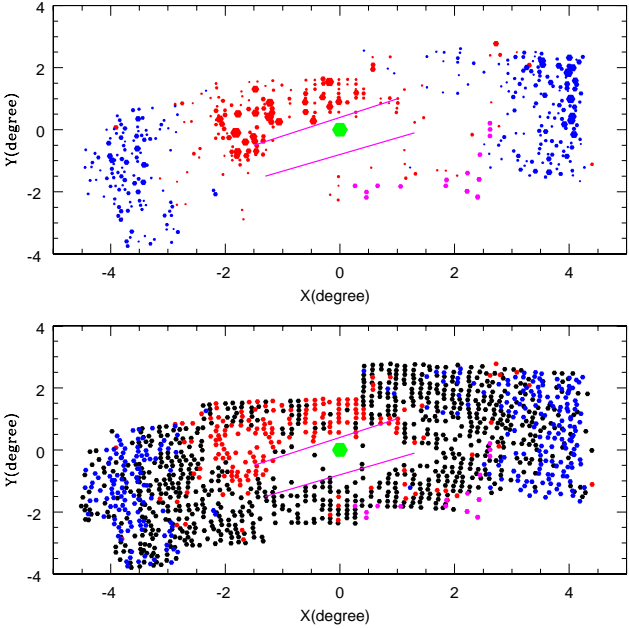}}
\caption{In the lower panel, black dots represent the regions on the fitted LMC
plane, red dots represent the regions behind the fitted plane and 
the blue dots represent the regions which are in front of the fitted 
plane. The upper panel shows only the regions with deviations, greater than 
3 sigma, from the LMC disk plane. The size of the points are proportional 
to the amplitude of the deviations. Magenta dots in both panels are 
the regions which are suggested as warps by \cite{os02}.
The green hexagon in both plots represents the optical center of the 
LMC.}
\end{figure}

The extra-planar features seen in the LMC disk would have affected 
the estimate of the planar parameters of the LMC disk. The reduced  
$\chi$$^2$ value for the estimate of planar parameters using the MCPS data set is 
2.0. The higher value of reduced $\chi$$^2$ value can be due to the presence 
of structures in the LMC disk. So we omitted the regions with deviations 
above 3 sigma and the planar parameters are estimated again. Around 320 
regions out of 780 regions showed deviations above 3 sigma and the remaining 
460 regions are used for the re-estimate. 
Thus the  structural parameters obtained for the 
LMC disk plane after removing the extra planar features are 
inclination, $i$ = 37$^o$.2$\pm$2$^o$.3 and PA$_{lon}$, $\phi$ = 
141$^o$.4$\pm$3$^o$.7. The reduced $\chi$$^2$ value for the plane fitting 
in this case is 0.4.\\   

The plane-fitting procedure applied to 1262 regions of the   
OGLE III data gives an $i$ = 21$^o$.0 $\pm$ 2$^o$.2 \& 
$\phi$ = 171$^o$.1 $\pm$ 8$^o$.8 for the LMC disk. As done for the 
MCPS data, deviations from the LMC disk plane with the above planar parameters 
are estimated. The average error in the estimate of z distance for the OGLE III 
data set is around 300 pc. The regions which show deviations above 3 sigma 
are considered as real deviations. Fig.3 shows the deviation plot for the 
OGLE III data. The bottom panel shows all the 1262 regions used for 
plane fitting and the upper panel shows only those regions which 
show deviations above 3 sigma. In the upper panel the size of the points are 
proportional to the amplitude of the deviations. The color code in the 
figure is the same as in Fig.2. As in the MCPS data, the OGLE III also shows, 
similar brightening in the southeast, northwest and southwestern regions 
of the LMC disk. The northeastern part adjacent to the LMC bar is also dimmer 
in this plot similar to the MCPS plot. So either the northwestern, south
western and southeastern parts of the LMC disk regions are in front 
of the fitted plane and/or the RC population in these regions is different. 
Again the northeastern part north of the LMC bar is either behind   
and/or the RC population in this region is different. As done for 
the MCPS data set, the extra-planar features are removed and 
the parameters are estimated again for the OGLE III data set.  
Out of 1262 regions, 397 regions showed deviations above 3 sigma, and these 
regions are removed from the plane-fitting procedure.  Thus the  
structural parameters obtained for the LMC disk plane after removing 
the extra planar features are inclination, $i$ = 23$^o$$\pm$0$^o$.8 
and $\phi$ = 163$^o$.7$\pm$1$^o$.5. The reduced $\chi$$^2$  
value of the plane-fitting procedure with all the 1262 regions was 1.0 and 
now, when the region with large deviations are removed, the reduced 
$\chi$$^2$ became 0.3.\\


After removing the extra-planar features, $\phi$ estimated 
for the MCPS and the OGLE III data sets are 141$^o$.2$\pm$3$^o$.7 
and 163$^o$.7$\pm$1$^o$.6 respectively.The dereddened RC magnitude 
is plotted against the axis perpendicular to the line of nodes, 
axis of maximum gradient and its shown in Fig.4. 
 The upper panel shows the plot for the MCPS data and the lower 
panel shows the plot for the OGLE III data. 
From both plots we can clearly see the effect of inclination.
The slopes estimated for both data sets excluding 
the regions with deviations above 3 sigma are 0.0254 mag/degree, and
 the y-intercept is 18.18 mag for the MCPS data and a slope of 0.0152 mag/degree 
and the y-intercept of 18.13 mag for the OGLE III. The inclination estimated 
for the MCPS data is 33.3 degrees and 21.5 degrees for the OGLE III. The 
inclinations estimated from the plots for both data sets match well 
with the inclination values estimated from the plane-fitting 
procedure for the same data set.\\

The y-intercepts obtained from Fig.4 for both data sets are  
the mean $I_0$ value of RC stars. As RC stars are standard 
candles, this value is a measure of the distance to the center of the LMC. 
The y-intercepts obtained for the OGLE III and the MCPS data sets from Fig.4 are 
18.13$\pm$0.01mag and 18.18$\pm$0.01 mag respectively. The distance modulus, $\mu_0$ to the 
LMC center can be estimated using the formula\\

$\mu_0$ = $I_0$$_{mean}$ - $M_I$$_{(LMC)}$.\\

In the above equation, $M_I$$_{(LMC)}$ is the I-band absolute magnitude 
of RC stars in the LMC. Stanek et al (1998) used the absolute I-band 
magnitude of RC stars in the Hipparcos sample as the zero point for 
distance estimation to the LMC. Later \cite{GS01} found from their simulations of local clump and those found in the LMC that 
there is a systematic magnitude difference between them. 
This systematic magnitude difference is due to the differences in the age, 
metallicity, and star-formation rate of the RC stars in the Galaxy and the LMC.
\cite{GS01} simulated the RC stars in the LMC bar 
as well as in the outer fields using the star-formation rate estimated by 
\cite{HGC99} and the age-metallicity relation from \cite{PT98}.
 From their simulations, they 
estimated a $M_I$$_{(LMC)}$ of $-$0.371 for the central region of the LMC bar. 
Thus the distance moduli, $\mu_0$ to the LMC center estimated from the
OGLE III and the MCPS data are 18.50$\pm$0.01 and 18.55$\pm$0.01 respectively after correcting for population effects. These values agrees well  
with the previous estimates of 18.5 $\pm$ 0.02 (\cite{A04})
and 18.53 $\pm$ 0.07 (\cite{SG02}) toward the LMC.\\

\begin{table*}
\centering
\caption{Summary of orientation measurements of LMC disk plane}
\label{Table:1}
\vspace{0.25cm}
\begin{tabular}{lrrrr}
\hline \\
Reference & Inclination, $i$  & PA$_{lon}$, $\phi$  
& Tracer used for the estimate\\ \\
\hline
\hline \\

\cite{df72} & 27$^o$$\pm$2$^o$ & 170$^o$$\pm$5 & Isophotes \\
\cite{F77} & 33$^o$.0$\pm$3$^o$ & 168$^o$$\pm$4$^o$ & HI\\
\cite{cc86} & 28$^o$.0$\pm$5$^o$.9 & 142$^o$.4 $\pm$7$^o$.7 & Cepheids\\ 
\cite{LR92} & $-$ & 162$^o$.0 & HI\\
\cite{k98} & 22$^o$.0$\pm$6$^o$ & 168$^o$.0 & HI \\
\cite{vc01} & 34$^o$.7$\pm$6$^o$.2 & 122$^o$.5$\pm$8$^o$.3 & AGB stars \\ 
\cite{os02} & 35$^o$.8$\pm$2$^o$.4 &145$^o$$\pm$4$^o$ & Red clump stars \\
\cite{n04} & 30$^o$.7$\pm$1$^o$.1 & 151$^o$$\pm$2$^o$.4 & Cepheids\\
\cite{p04} & 27$^o$.0$\pm$6$^o$.0 & 127$^o$$\pm$10$^o$.0 & Cepheids\\
\cite{k09} & 23$^o$.5$\pm$0$^o$.4 & 154$^o$.6$\pm$1$^o$.2 & Red clump stars \\ \\
\hline
\hline \\
Our estimates\\ 
\hline\\
OGLE III & 23$^o$$\pm$0$^o$.8 & 163$^o$.7$\pm$1$^o$.5 & Red clump stars\\
MCPS  & 37$^o$.4$\pm$2$^o$.3 & 141$^o$.2$\pm$3$^o$.7 & Red clump stars\\\\
\hline
\end{tabular}
\end{table*}


\section{Discussion}
The dereddened peak I magnitude of RC stars from the 
OGLE III and the MCPS data sets are used to estimate the structural
parameters of the LMC disk and hence the deviations of 
the LMC regions from the plane. The planar parameters of the 
LMC disk obtained from the analysis of 1262 regions of the
OGLE III data turned out to be an inclination of $i$=23$^o$$\pm$0$^o$.9 
and PA$_{lon}$, $\phi$ = 163 $^o$.7$\pm$1$^o$.6. From the analysis of 
780 regions of the MCPS data, an inclination, $i$=37$^o$.4$\pm$2$^o$.5 and PA$_{lon}$
$\phi$ = 141$^o$.2$\pm$3$^o$.7 are obtained. Previously many studies have 
been done to obtain the planar parameters of the LMC disk using various 
tracers. The values obtained from those studies along 
with our estimates are summarised in Table 1. Tracers used in those 
studies are also mentioned in the table. Our estimates based on the 
analysis of the MCPS data is matching well within the error bars with 
the estimates of \cite{os02}. Various studies of the LMC disk 
and bar regions (Fig.6 given in \cite{k09}, Fig.2 in \cite{ss09apj} and 
Fig.4 in \cite{s03}) have shown that it is a highly structured galaxy. The 
estimate of the planar parameters are likely to be severely 
affected by these structures. 
The difference in the regions, coverage of the LMC, and the tracer used 
for the estimate of the planar parameters may be the reason for the  
differences in the estimated parameters of various studies. 
Also, \cite{n04} showed that 
the analysis based on the photometric data from the concentric rings in the 
inner LMC  is strongly dependent on the adopted LMC center, which can cause 
a variation of about 35 degrees in the values of PA$_{lon}$.\\

\begin{figure}
\resizebox{\hsize}{!}{\includegraphics{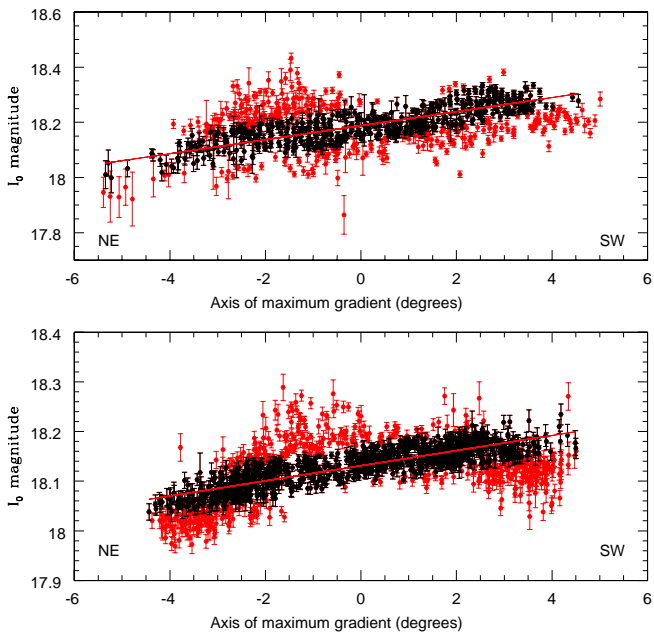}}
\caption{Dereddened RC magnitude plotted against the axis of maximum   
gradient.  The red points are regions which show deviation larger than 3 sigma.
 The direction of inclination is shown as red line. The upper panel is for 
the MCPS data and the lower panel is for the OGLE III.}
\end{figure}

\subsection{Choice of the LMC center}
The adopted center of the LMC for the analysis 
may have an effect on the estimated parameters.
We used the plane fitting method, rather than the ring analysis, 
which is probably less affected by the choice of the center. 
For our analysis, we have taken the optical center, $05^h19^m38^s.0$ 
$-69^o27'5".2$ (J2000) (\cite{df72}), as the 
center of the LMC. To see the effect of the adopted center on 
the estimated parameters, we did the analysis with the 
HI rotation center (\cite{k98}), geometric center of the sample Cepheids 
used in the analysis of \cite{n04} and also the carbon stars isopleths 
center (\cite{vc01}) as the center of the LMC. 
We find that the choice of center causes only marginal changes in the estimated parameters,
and the changes are within the error. The estimated parameters with different 
centers are given in Table 2.\\

\begin{table*}
\centering
\caption{Summary of orientation measurements of the LMC disk plane with choice of the
center}
\label{Table:2}
\vspace{0.25cm}
\begin{tabular}{lrrr}
\hline \\
Reference & Inclination, $i$  & PA$_{lon}$, $\phi$  
\\ \\
\hline
\hline \\

MCPS data\\

Optical center \cite{df72}& 37$^o$.4$\pm$2$^o$.3 & 141$^o$.2$\pm$3$^o$.7 \\
HI rotation center (\cite{k98})& 37$^o$.2$\pm$2$^o$.2 & 142$^o$.7$\pm$3$^o$.4 \\
Geometric center of the sample Cepheids used in the analysis of \cite{n04}& 36$^o$.8$\pm$2$^o$.2 & 141$^o$.4$\pm$3$^o$.8 \\
Center of carbon stars outer isopleths (\cite{vc01})& 36$^o$.9$\pm$2$^o$.5 & 137$^o$.4$\pm$4$^o$.0 \\
\hline
\hline \\

OGLE III data\\ 

Optical center \cite{df72}& 23$^o$.0$\pm$0$^o$.8 & 163$^o$.7$\pm$1$^o$.5 \\
HI rotation center (\cite{k98})& 23$^o$.1$\pm$0$^o$.8 & 164$^o$.9$\pm$1$^o$.7 \\
Geometric center of the sample Cepheids used in the analysis of \cite{n04}& 22$^o$.7$\pm$0$^o$.8 & 164$^o$.0$\pm$1$^o$.6 \\
Center of carbon stars outer isopleths (\cite{vc01})& 22$^o$.0$\pm$0$^o$.8 & 161$^o$.9$\pm$1$^o$.6 \\
\hline
\hline \\
\end{tabular}
\end{table*}

\subsection{Choice of area of the regions}
In order to check the effect of area of the bins used in our analysis, the 
OGLE III region is divided into sub regions with area, 8.88 x 16.76 arcmin$^2$,
comparable to the area of sub-region of the MCPS data. The whole analysis is 
repeated and the structural parameters of the LMC 
disk are estimated. The parameters, inclination and PA$_{lon}$ 
obtained are 24.5$^o$$\pm$0$^o$.5 \& 166$^o$$\pm$1$^o$  respectively. 
These values are comparable with our results obtained from the 
analysis of the OGLE III data with smaller area (8.88 x 8.88 arcmin$^2$) 
sub-regions. This means that extending the area of sub-regions is unlikely to change 
the estimated parameters. But when the area of the OGLE III sub-region 
is made 4.44 x 4.44 arcmin$^2$, smaller than 8.88 x 8.88 arcmin$^2$, 
used for our analysis, there was problem in the plane-fitting procedure 
due to the large number of points. 
 This can be due to the finer structures present in the inner LMC.\\

\subsection{Effect of reddening}
\begin{figure}
\resizebox{\hsize}{!}{\includegraphics[width=12cm]{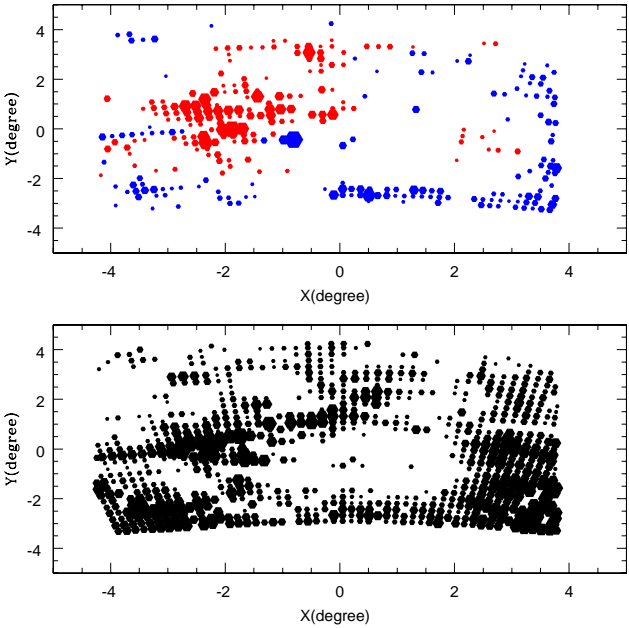}}
\caption{Two-dimensional plot of the extinction, A$_I$ for the MCPS data 
is shown in the lower panel and a two-dimensional plot of deviation is shown 
in the upper panel. In the lower and upper panels the size of the points 
are proportional to the amplitude of the extinction and deviation
of the regions from the plane of the LMC disk respectively. The blue and red 
points in the upper panel are the regions which are in front 
 and behind the fitted plane respectively.}
\end{figure}

\begin{figure}
\resizebox{\hsize}{!}{\includegraphics[width=12cm]{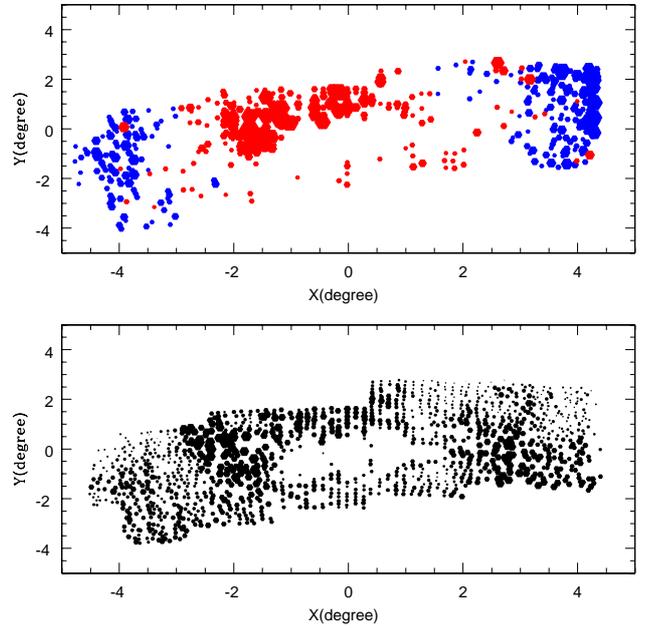}}
\caption{Two-dimensional plot of the extinction, A$_I$ for the OGLE III data 
is shown in the lower panel and a two-dimensional plot of deviation is shown 
in the upper panel. In the lower and upper panels the size of the points 
are proportional to the amplitude of the extinction and deviation
of the regions from the plane of the LMC disk respectively. The blue and red 
points in the upper panel are the regions which are in front 
 and behind the fitted plane respectively.}
\end{figure}
The other important point to be discussed is the role of reddening in the 
estimate of the structure of the LMC disk. The extra-planar features which 
are found both behind the disk and in front of the disk could be in the plane 
of the LMC disk itself if there were an over-estimate or under-estimate 
of the reddening. 
It has been demonstrated by \cite{Z97} that the 
extinction property of the LMC varies both spatially and as a function 
of stellar population. In our study, the dereddening of RC stars is done 
using the reddening values estimated from the RC stars itself.
  
But again, to understand the effect of reddening on the detected 
extra-planar features, we plotted a two dimensional plot of reddening as well 
as the deviations. The plots for the MCPS and the OGLE III data are shown in Figs.5 
and 6 respectively. In both the plots the lower panel shows the  
reddening distribution and the upper panel shows the distribution  
of the deviations. In the lower panel the size of the point is proportional 
to the reddening value and in the upper plot the size of the point is 
proportional to the amplitude of the deviation. The red points in the upper 
panel of the plot represents the regions behind the plane and blue points 
represent the regions in front of the plane. We can see that in the MCPS plot 
(Fig.5) the regions in the southwestern part of the LMC disk around our 
suggested warps show more reddening. Hence it is possible that these warps are due to 
over-estimate of reddening in these regions. \cite{os02} also 
found that the reddening near the southwestern part of the disk 
near the regions of their suggested warps is stronger. They correlated 
the large reddening in the southwestern LMC regions to the 
diffuse extinction due to the high Galactic foreground dust, 
having A$_I$ of approximately 0.3 mag,  given in the COBE-DIRBE-IRAS/ISSA 
dust map (\cite{sch98}). \cite{os02} measured an A$_I$ of 
approximately 0.25 mag at the LMC's southwest edge. We measured an 
A$_I$ ranging from 0.1 to 0.2 with an average of 0.16 mag 
at the LMC southwest edge. The plot which shows the amplitude of the deviation 
against the extinction values for the MCPS region is shown in Fig.7. 
The magenta dots are the regions in the southwestern region of LMC, which 
are brighter than the surrounding regions. 
The reddening is high in these regions compared to the surrounding regions.
There are also regions which are in front of the plane, like the southeastern disk, 
which do not show large reddening.
Again, regions in the northeastern part with respect to the LMC bar show large 
reddening and these regions are shown in the deviation plot as behind the 
fitted plane.  Subramaniam et al(2010) (in preparation) show that these regions 
coincide with the star-forming regions. Once again, it could be argued that the
reddening has not been accounted for properly here. There are also some nearby regions
behind the fitted plane, which do not show large reddening. Thus some regions which are in the front
or behind  show a range of reddening as seen in Fig.7. On the other hand,
we do not see a strong correlation between reddening and the deviation, because both
positive and negative deviations are observed for regions with large reddening.
That is, the reddening could not have been both under and over-estimated.
Thus, reddening
is not strongly correlated with the estimated structures, but the extend of the
deviation from the LMC plane may still be affected by the reddening that is present.\\

\begin{figure}
\resizebox{\hsize}{!}{\includegraphics[width=12cm]{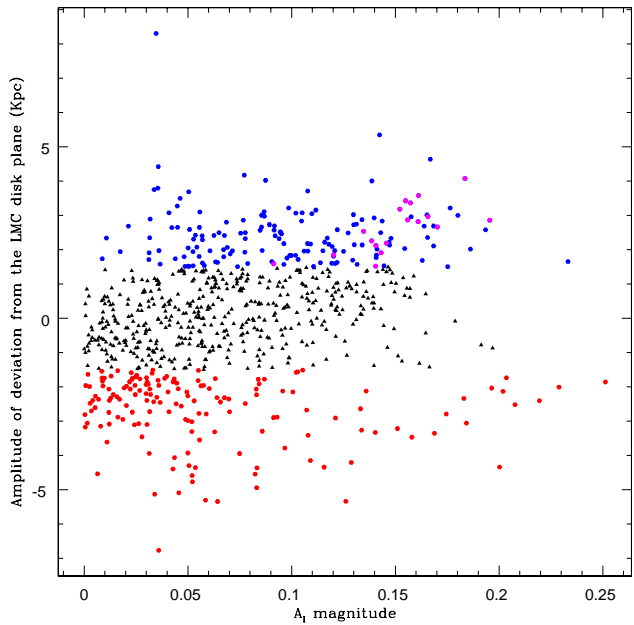}}
\caption{Amplitude of the deviations of the regions from the fitted plane are plotted against the A$_I$ mag for the 
 MCPS data. Black points are those with 
deviations less than 3 sigma, and red and blue are those with deviations 
more than 3 sigma and are behind the plane and in front of the plane 
respectively. Magenta dots are regions which are in front of the fitted plane 
in the southwestern part of the LMC disk.}   
\end{figure}
 
As mentioned in the analysis section, while estimating the reddening values 
for the MCPS regions, 451 regions out of 1231 regions showed a negative 
value for the reddening. We removed those regions from our analysis. 
 \cite{z04} estimated the extinction map for 
the LMC by comparing the stellar atmospheric models and observed 
colors using the MCPS data. They also found that many regions show 
negative extinction values. They suggested that large observational uncertainties 
scatters the extinction to negative values, and they set those negative 
values as zero extinction values. We estimated the structural 
parameters of the LMC disk using all the 1231 regions of the MCPS data after 
assigning zero reddening value for those regions which show negative value 
for the reddening. A weighted plane-fitting procedure was applied to the  
1231 regions of the MCPS data and deviations were estimated.
Deviations above 3 sigma were considered as deviations, and after removing 
those regions, planar parameters of the LMC disk plane was re-estimated. An 
inclination, $i$ = 39$^o$.3$\pm$3$^o$.2 and PA$_{lon}$ $\phi$ = 139$^o$.0$\pm$4 
was obtained. The planar parameters are matching well with the parameters
 we estimated without considering the regions which showed negative values 
for reddening. 
To find where these regions are located with respect to the plane of the LMC disk, 
we plotted the reddening E(V-I) mag vs amplitude of the deviation calculated. 
This plot is shown in Fig.8. We can clearly see that most of the regions which are assigned 
zero reddening values show large deviation from the plane. These 
regions are found both in front and behind the fitted plane. These 
deviations may be real. Because the reddening has an important role 
in the estimation of the structural parameters of the disk and the 
amplitude of the deviations, we prefer the analysis without considering regions with negative 
reddening. The agreement in the estimated parameters could just be 
coincidence because the deviations estimated are both in front  
and behind the plane and thus the net effect on the parameters of the LMC plane
is minimum.\\

\begin{figure}
\resizebox{\hsize}{!}{\includegraphics[width=12cm]{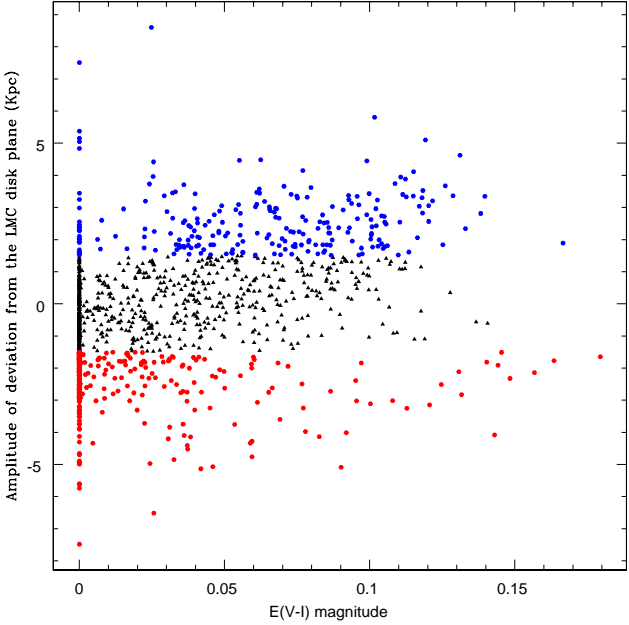}}
\caption{Amplitude of the deviations of the regions from the plane of the 
LMC disk plane are plotted against the reddening, E(V$-$I) values for the 
 MCPS data after assigning zero reddening for the regions 
which show negative values for the reddening. 
(see Fig.7 for details)}   
\end{figure}
   
For the MCPS data in general the color of the RC stars is bluer when
compared to the OGLE III RC stars. 
This can be due to the difference in the filter systems
used and/or calibrations. The choice of intrinsic (V$-$I) color of 
RC stars can be another reason for obtaining negative reddening values 
for a large number of regions in the MCPS data. The intrinsic color of RC 
stars for calculating reddening is taken to be 0.92 mag (\cite{os02}). 
\cite{os02} selected this value to produce the median reddening obtained 
toward the LMC by \cite{sch98}. For the OGLE III data set we calculated the 
intrinsic (V$-$I) color of RC stars to produce the median reddening 
obtained by \cite{sch98}. The value turned out to be 0.915 mag, which
is similar to the value given by \cite{os02}. So we did the 
analysis of the OGLE III data with intrinsic color value, 0.92 mag. 
Also, there are no locations in the OGLE III data, where the value of 
estimated reddening goes negative.
We calculated the intrinsic (V$-$I) color of
RC stars in the MCPS data set in a similar way as \cite{os02}. The
value obtained is 0.84 mag. Using this value we repeated the analysis
explained in Sect.3 to estimate the LMC disk parameters. Now, out of 1231
locations, only 64 regions showed negative value for reddening. 
A plane fitting procedure is applied to 1167 regions 
and deviations from the plane are also estimated. 
The regions with deviations above 3 sigma are removed and the plane 
fitting procedure is applied to the remaining regions. The values 
obtained are $i$= 38$^o$.7$\pm$2$^o$.4 and $\phi$ = 150$^o$.9$\pm$3$^o$.8.   
The inclination value is comparable and within errors with the 
analysis done with intrinsic (V$-$I) color of RC stars, 0.92 mag. 
But the PA$_{lon}$ value is higher than those obtained in the earlier analysis.
Thus by including more regions in the inner LMC, the PA$_{lon}$ increased
significantly, whereas the inclination did not change.\\

\subsection{Effect of coverage and inner structure of the LMC disk}
\begin{table*}
\centering
\caption{Summary of orientation measurements of the LMC disk plane at different radii}
\label{Table:3}
\vspace{0.25cm}
\begin{tabular}{lrrr}
\hline \\
Region & Inclination, $i$  & PA$_{lon}$, $\phi$  
\\ \\
\hline
\hline \\

MCPS data\\

r $>$ 1 & 36$^o$.1$\pm$2$^o$.2 & 141$^o$.8$\pm$3$^o$.9 \\
r $>$ 2 & 34$^o$.2$\pm$1$^o$.7 & 143$^o$.3$\pm$4$^o$.3 \\
r $>$ 3 & 29$^o$.8$\pm$1$^o$.1 & 151$^o$.7$\pm$5$^o$.0 \\
r $>$ 4 & 39$^o$.4$\pm$2$^o$.4 & 147$^o$.8$\pm$1$^o$.5 \\
\hline
\hline \\

OGLE III data\\ 

r $>$ 1 & 23$^o$.5$\pm$0$^o$.8 & 162$^o$.0$\pm$1$^o$.6 \\
r $>$ 2 & 23$^o$.2$\pm$0$^o$.5 & 160$^o$.6$\pm$1$^o$.3 \\
r $>$ 3 & 28$^o$.4$\pm$0$^o$.2 & 146$^o$.9$\pm$1$^o$.9 \\
r $>$ 4 & 35$^o$.6$\pm$1$^o$.7 & 137$^o$.8$\pm$2$^o$.0 \\
r $\le$ 3& 16$^o$.5$\pm$0$^o$.9 & 161$^o$$\pm$4$^o$.5 \\ 
\hline
\hline \\
\end{tabular}
\end{table*} 
The estimated planar parameters of the LMC may vary depending on  
the difference in the coverage of the LMC studied to estimate these parameters. 
We also obtained significantly different results from the MCPS and the 
OGLE III data. In order 
to understand the effect due to coverage, we estimated the planar parameters of the LMC disk 
excluding the inner LMC data at different radii from the LMC center. Thus the planar 
parameters are estimated using the data above the radius of 1, 2, 3, and 4 degrees from the 
LMC center. The results are summarised in Table.3. 
One of the important results of this analysis is that   
the parameters estimated excluding the data within the radius of 3 
degrees from the LMC center in both the MCPS and the OGLE III sets are similar,
as shown in Table.2. 
From Table.1 we can see that there is a significant difference in the 
estimates of the planar parameters, from the analysis of full data set of the
OGLE III and the MCPS data. When the 
inner data are excluded from the analysis, the results are matching within the error bars. 
This indicates that the inner structure of the LMC disk  affects the 
estimate of the LMC disk parameters. 
Even though both the MCPS and the OGLE III contain data within 
3 degrees of the radius from the LMC center, the OGLE III data are more affected by the inner 
structures.
This is because the OGLE III contains a highest number of 
inner regions compared to the MCPS data. As mentioned before, several  
central regions had to be
removed from the analysis of the MCPS data because they showed negative reddening. 
 That probably is the reason for a high PA$_{lon}$ and low inclination obtained 
from the OGLE III analysis.\\

To understand the inner structure of the 
LMC, we tried fitting a plane to the data within 3 degrees from the LMC center. 
For the MCPS data the fitting was a problem because of fewer points and also because of the 
deviations present within the 3 degree region from the center. For the OGLE III data the 
parameters obtained are $i$= 16$^o$.5$\pm$0$^o$.9 and PA$_{lon}$ = 161$^o$$\pm$4$^o$.5. 
This indicates that the inner 
and outer structures of the LMC disk are different. Regions within the 3 degree radius fit to
a plane with significantly less inclination and large PA$_{lon}$, whereas regions outside 3 degrees
have a large inclination and less PA$_{lon}$.\\

A similar analysis was performed by \cite{vc01}, for AGB stars, where they
estimated the variation of $i$ and PA$_{lon}$ as a function of radius.
They used rings, whereas we used concentric regions progressively excluding the inner regions. 
To compare our results with theirs, we also obtained parameters of 
the LMC disk in different rings, and the results are given in Table.4.
They found that the $i$ and PA$_{lon}$ decrease with radius, whereas we find only the 
PA$_{lon}$ to decrease with radius. 
Thus both analyses agree that the PA$_{lon}$ decreases with increasing radius.
 The differences in the trend seen for the inclination may be due to the difference in the tracer and the method used. 
The inclination $i$ is found to be increasing 
with radius for the OGLE III data. For the MCPS data, we find that the outer two rings show 
an increasing $i$, whereas the innermost ring does not follow the trend. Therefore a definite pattern is not seen for the MCPS data set. 
Combining the result obtained in the last paragraph, we suggest that there is a difference in the 
structure of the LMC inside and outside a radius of around 3 degrees. Our results suggest that
the outer disk is inclined more than the inner disk with a reduced PA$_{lon}$.
\cite{sapjl05} 
proposed the existence of two disks in the inner 3$^o$ of the LMC, 
with one counter rotating. May be the existence of two disks  
in the inner LMC is the reason that it is different from the outer structure.\\
\\

\begin{table*}
\centering
\caption{Summary of orientation measurements of the LMC disk plane at different radial rings}
\label{Table:4}
\vspace{0.25cm}
\begin{tabular}{lrrr}
\hline \\
Region & Inclination, $i$  & PA$_{lon}$, $\phi$  
\\ \\
\hline
\hline \\

MCPS data\\

2.5$<$r$<$3.5 & 38$^o$.9$\pm$0$^o$.3 & 161$^o$.9$\pm$3$^o$.5 \\
3.5$<$r$<$4.5 & 25$^o$.2$\pm$1$^o$.4 & 151$^o$.4$\pm$7$^o$.5 \\
4.5$<$r$<$5.5 & 37$^o$.7$\pm$1$^o$.4 & 134$^o$.2$\pm$0$^o$.7 \\
\hline
\hline \\

OGLE III data\\ 

2.5$<$r$<$3.5 & 21$^o$.4$\pm$0$^o$.3 & 157$^o$.6$\pm$2$^o$.6 \\
3.5$<$r$<$4.5 & 31$^o$.4$\pm$0$^o$.4 & 139$^o$.5$\pm$1$^o$.4 \\
4.5$<$r$<$5.5 & 39$^o$.1$\pm$1$^o$.6 & 134$^o$.6$\pm$0$^o$.9 \\
\hline
\hline \\
\end{tabular}
\end{table*}

\subsection{Deviations and warps}
After fitting the LMC plane, the deviations of the disk with respect
to the fitted plane seen in the overlapping regions of the OGLE III and the MCPS 
data sets are similar. The extra-planar feature, which is seen as 
behind the LMC plane in the northeast of the LMC bar is interesting. 
This feature is seen in both the MCPS and the OGLE III deviation plots. 
An extension behind the disk in the same region is suggested by 
\cite{ss09aal} in the study of RR Lyrae stars in the LMC. 
Some of the star-forming regions (\cite{k00})  are lying around 
this feature. 
Brightening of the RC stars in the western and eastern ends of the disk are 
seen in the OGLE III deviation plot. The brightening in the northwestern part 
of the LMC disk is more clearly seen with larger amplitude. The deviation 
of the the MCPS data shows a similar brightening in the northwestern, southwestern,
 and southeastern regions of the LMC disk. The warps suggested 
by \cite{os02} in the southwestern region of the LMC are overplotted 
in our deviation plots, and not many regions which they suggest as 
warps are coinciding with our deviations, though they are located 
near the regions where we see deviations in the 
southwestern disk. Thus the brightening of the RC stars in the 
southwestern part of the LMC disk and the dimming of RC stars in the 
northeastern part near the LMC bar are suggestive of a 
symmetric warp in the LMC disk. \cite{n04} suggested a 
symmetric warp in the LMC disk similar to our results. 
Along with the symmetric warp, the brightening of RC stars in the northwestern 
part of the LMC disk is also very clearly seen in both the MCPS and the OGLE III 
deviation plots. The southeastern part of the LMC disk is also comparatively 
brighter than other regions, which makes the warp asymmetric.
In the the MCPS deviation plot, brightening of RC stars is 
seen near the optical center,  in the bar region of the LMC. This kind of 
RC brightening is seen near the optical center in the bar region of the LMC 
by \cite{ss09apj} and also by \cite{k09}.\\ 

\begin{figure}
\resizebox{\hsize}{!}{\includegraphics[width=12cm]{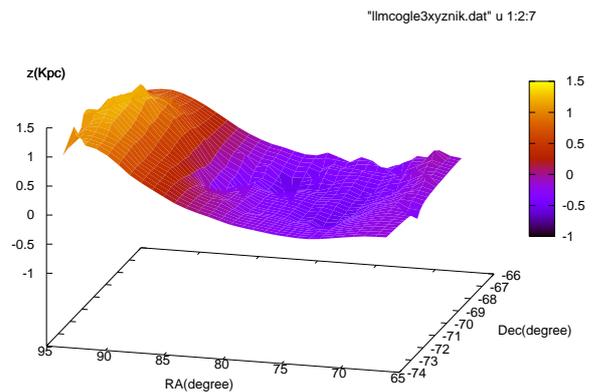}}
\caption{3D plot of the LMC disk obtained from the OGLE III RC stars.
}   
\end{figure}
   
The edge-on view of the LMC as seen from the minor-axis (Fig.4) also suggests that the inner LMC
might have a different inclination with respect to the outer regions. Apart from this,
the symmetric warp can also be noticed in this plot. It can also be noticed that the warp is
not centered at the LMC center, but shifted towards the southwest by about a degree. Thus, 
our results suggest an off-centered symmetric warp. It is quite clear from the analysis that
the inner structure of the LMC is quite complicated. In order to get a 3-D picture of the region
studied here, we have shown the RA, Dec, and Z plot of the OGLE III region in Fig.9. The surface plot shown here 
is obtained with a standard surface plot algorithm and thus involves interpolation and averaging.
Therefore, this plot can be used to obtain a qualitative understanding of the structure, but cannot
be used for any quantitative analysis. The figure clearly shows that the inner regions have a smaller inclination and
the outer regions have large inclination. The increase in the inclination seems to start at a closer radius
in the northeast, when compared to the southwest, which makes it off-centered. It is clear from this figure
that the LMC disk shows a lot of structures, and it is difficult to define a plane. By fitting a plane,
a large number of regions are likely to show deviation from the plane suggesting warps. 
It is also clear that depending on the coverage of the data, the plane fitted and the parameters obtained
are likely to change, which in turn will identify different regions to have deviations/warps.
Also, defining a warp gets complicated due to contribution from reddening and population effects. 
In general, we notice that the inclination increases with radius. The northeast of the LMC is closer
to us, and we find that it gets even closer with radius, with an increase in inclination. 
The southwest part, instead of getting more distant,
also seems to get slightly closer. This is also seen as a warp by \cite{os02}. Thus, the plot
suggests that studies which ignored the inner regions and considered only the outer regions are likely
to derive a highly inclined plane with a smaller PA$_{lon}$. Studies that did not consider the outer
regions would derive a less inclined plane with a large PA$_{lon}$. Because the change in the inclination
is off-centered, methods which use ring-analysis are likely to be more affected by the inner structure.\\

Warps and structural changes seen in the LMC disk could be due to
tidal interactions. The Small Magellanic Cloud is unlikely to be the cause because it is smaller compared to the LMC.
This effect may be due to the gravitational attraction of our Galaxy on the LMC.
Thus, our results point in the direction that the increased inclination and the warps identified in the outer regions may be
due to the interaction with our Galaxy. 
It is important to study more distant regions to understand the change in the disk structure with radius.
A detailed study of the outer structure may throw light on the
details of the tidal effects on the LMC disk and its origin.\\

\subsection{Comparison with the structure of the HI gas disk}
It is interesting to compare the structure of the HI gas disk of the LMC 
with that of the stellar disk. Various HI studies have estimated the 
PA$_{lon}$ and the inclination of the HI disk by kinematical as well as 
by geometrical methods. These studies also suggested the presence of 
extra-planar features in the HI disk.
\cite{F77} derived an inclination of 33$^o$$\pm$3$^o$ and PA$_{lon}$, 168$^o$$\pm$4$^o$ 
by geometrical means. 
The HI velocity studies by \cite{LR92} revealed two 
kinematic components, the L (lower velocity) component and the D (disk) 
componenet. The D component was found to be extended in the whole LMC, 
and the L component, with two deformed lobes, was found to the north of 30 Dor 
and south of 30 Dor region. The L component was found to be around 50-500 pc 
above the D component. The PA$_{lon}$ of around 162$^o$ was 
estimated for the disk component by kinematical method (line of maximum 
velocity gradient). The above values are not very different from the PA$_{lon}$ of 
the stellar disk estimated from the OGLE III data.
\cite{k98} estimated the PA$_{lon}$ of HI disk
to be around 168$^o$ by kinematical method. They suggested that the HI disk
is inclined in a way that PA=78$^o$ is closer to us.
This value is agrees well with 
our result, which gives the PA of the closest part to be around 71$^o$ $\pm$ 
4$^o$.5 for the inner stellar disk from the OGLE III data. 
\cite{k98} also estimated the inclination of the HI disk 
from the outer isophotes of HI brightness temperature 
to be  around 22$^o$. As this value was poorly determined and highly
 correlated with other parameters, they adopted the canonical value of 33$^o$
(\cite{wes97}).\\

\begin{figure}
\resizebox{\hsize}{!}{\includegraphics[width=12cm]{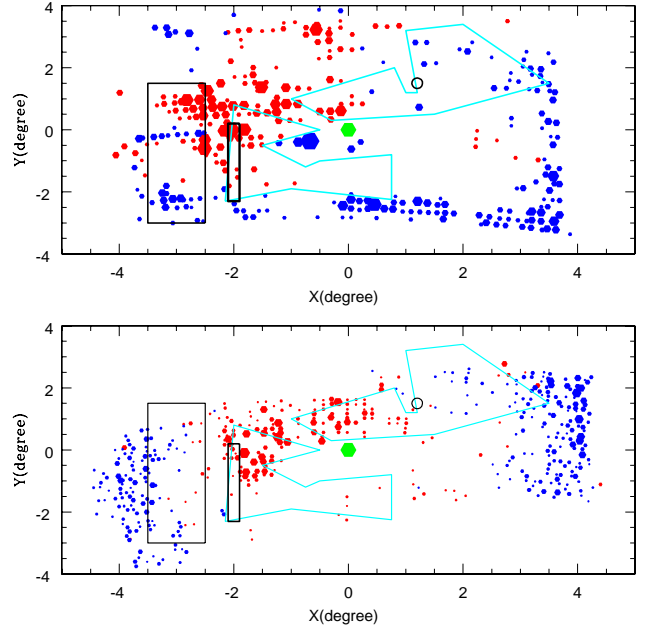}}
\caption{Deviations found in the stellar disk of the LMC 
using the OGLE III and the MCPS data are plotted in the lower and 
upper panels respectively. The color code is the same as in Fig.2 and Fig.3.
 The two cyan lobes in both the panels are the L component identified in 
HI disk by \cite{LR92}. The black features in both the panels are the 
kinematical warps suggested by \cite{LR92} in the L and D components.
}   
\end{figure}
In order to compare the HI and stellar disk structures, 
we plotted the deviations found in the stellar disk from the OGLE III and 
the MCPS data sets along with the structures found in the HI disk.
In Fig.10 the lower panel shows the deviations 
found in the OGLE III data and the upper panel shows the deviations found in 
the MCPS data set for the stellar disk. 
As in the previous plots, red points denote the regions behind the fitted plane 
and blue points denote the regions in front of the fitted plane, with sizes
proportional to the amplitude of the deviations.
The approximate location of the L component identified by 
\cite{LR92} is shown as two irregular structures in 
cyan color. \cite{LR92} suggested the presence of kinematical 
warps in some regions of the D component as well as 
in some regions of the L component. These regions are shown as 
black features.
The kinematical warps identified in the D component are in the big 
rectangular box in the east.  
The regions where kinematical warps are identified in the L component 
are shown as black circle and small black rectangular box. 
Most of the kinematical warps identified in the HI disk are more or less near 
the extra-planar features found in the stellar disk.\\ 

The PA$_{lon}$ estimated for the HI disk is similar to the PA$_{lon}$ of 
the stellar disk in the inner regions of the LMC, particularly the estimates
from the OGLE III data. The effect of 
inclination of the LMC, which makes the northeastern part appear to be 
closer, is seen both in stellar as well as in the HI disk 
of inner LMC. The inclinations are also similar.
The kinematic warp seen in the D component coincides with the eastern stellar warp.
There is a mild indication that the L component coincides with regions which
are located behind the disk. On the whole, the L component and the kinematical
warps identified in both components of HI, more or less coincide with the warps in the stellar disk.
There is no HI counter part for the southwestern warp found in the stellar disk.
These results suggest that the inner stellar and HI disk structures of the LMC 
are similar. \\

\section{Conclusions}
We used the RC stars identified from the OGLE III and the MCPS to estimate the structural parameters
of the LMC. The results can be summarised as follows:
\begin{description}
\item {We estimated the structural parameters of the
 LMC disk such as the inclination, $i$, and the position angle of the
 line of nodes (PA$_{lon}$), $\phi$ using a weighted least-square plane-fitting procedure.} 
\item {We find an inclination of $i$ =23$^o$.0$\pm$0$^o$.8 and PA$_{lon}$, $\phi$ =
 163$^o$.7$\pm$1$^o$.5 using 
the OGLE III data and an inclination of $i$=37$^o$.4$\pm$2$^o$.3 and PA$_{lon}$ 
$\phi$ = 141$^o$.2$\pm$3$^o$.7 using 
the MCPS data. Extra-planar features, which are in front 
as well as behind the fitted plane, are seen in both the data sets.}
\item {We find that the choice of center has a negligible effect on the estimated parameters.}
\item{The reddening is found to be large in some regions located in front and also in regions located
behind the fitted plane. Deviations are also found in regions without large reddening. 
These suggest that the reddening is only mildly correlated with
the deviations of the disk from the fitted plane.}
\item{We find that the disk within a 3 degree radius has 
lower inclination and higher PA$_{lon}$, and differs from the outer disk.} 
\item{The edge-on view of the LMC disk along the minor axis suggests an off-centered
symmetric warp.}
\item{The change of structure in the outer LMC could be due to tidal effects.}
\item{ We suggest that the complicated structure of the inner LMC causes variation
in the estimated planar parameters depending on the area covered for each study, including the
two data sets used here.}
\item{In the inner LMC, the stellar as well as the HI disk have similar properties.}
\end{description}
\acknowledgements
Smitha Subramanian acknowledges the financial support provided by Council of Scientific and Industrial Research (CSIR) , India through SRF grant, 09/890(0002)/2007-EMR-I. The authors thank the OGLE and the MCPS team for making the data available in public. We thank Knut Olsen for a critical reading of the manuscript and comments. We thank the referee for
suggestions which improved the paper.

\end{document}